\documentclass[%
 reprint,
 groupedaddress,
 amsmath,amssymb,
 aps,
 pra,
 longbibliography
]{revtex4-2}

\usepackage{graphicx}
\usepackage{subfigure}
\usepackage{xcolor}
\usepackage{fixmath}
\usepackage{microtype}
\usepackage{float}

\begin{document}

\title{Soft-Quantum Algorithms}

\author{Basil~Kyriacou}
\author{Mo~Kordzanganeh}
\author{Maniraman~Periyasamy}
\author{Alexey~Melnikov}
\affiliation{Terra Quantum AG, 9000 St.~Gallen, Switzerland}

\begin{abstract}
    Quantum operations on pure states can be fully represented by unitary matrices. Variational quantum circuits, also known as quantum neural networks, embed data and trainable parameters into gate-based operations and optimize the parameters via gradient descent. The high cost of training and low fidelity of current quantum devices, however, restricts much of quantum machine learning to classical simulation. For few-qubit problems with large datasets, training the matrix elements directly, as is done with weight matrices in classical neural networks, can be faster than decomposing data and parameters into gates. We propose a method that trains matrices directly while maintaining unitarity through a single regularization term added to the loss function. A second training step, circuit alignment, then recovers a gate-based architecture from the resulting soft-unitary. On a five-qubit supervised classification task with 1000 datapoints, this two-step process produces a trained variational circuit in under four minutes, compared to over two hours for direct circuit training, while achieving lower binary cross-entropy loss. In a second experiment, soft-unitaries are embedded in a hybrid quantum-classical network for a reinforcement learning cartpole task, where the hybrid agent outperforms a purely classical baseline of comparable size.
\end{abstract}

\maketitle

%% ============================================================
%%  INTRODUCTION
%% ============================================================
\section{Introduction}\label{sec: Intro}

    Variational quantum circuits (VQCs)~\cite{peruzzo2014variational, mitarai2018quantum}, sometimes called quantum neural networks~\cite{benedetti2019parameterized}, are a class of quantum algorithms that rely on classical optimization. They have gained attention for applications in quantum machine learning and quantum chemistry~\cite{sagingalieva2023hybrid}, and now form the backbone of most near-term quantum algorithms~\cite{cerezo2021variational, qml_review_2023}. The output of a VQC is

    \begin{equation}\label{eq:vqc}
        f(\mathbold{x, \theta}) = \text{tr}[ O(\mathbold{x, \theta}) \rho (\mathbold{x, \theta})]
    \end{equation}

    where $\rho$ is a density matrix and $O$ is a Hermitian observable. Both the density matrix and the observable can depend on inputs and trainable parameters. If $\rho$ defines a pure state starting in $|0 \rangle \langle 0 |$, it takes the form

    \begin{equation}
        \rho (\mathbold{x, \theta}) = U(\mathbold{x, \theta}) |0 \rangle \langle 0 |U^\dagger(\mathbold{x, \theta}),
    \end{equation}

    where the unitary operation $U(\mathbold{x, \theta})$ represents the circuit's cumulative product of gates. Typically, a VQC is defined by a set of gates, some of which hold trainable parameters that are optimized by a classical computer against a given loss function. In quantum machine learning, for instance, a portion of the total gates encodes the data while the parameters of the remaining gates are trained.

    Due to the high cost and low fidelity of quantum hardware~\cite{preskill2018quantum, bharti2022noisy}, NISQ-era VQCs are mostly trained on quantum simulators, that is, classical hardware that simulates quantum devices. Unlike classical data science, simulating VQCs struggles to scale: training times grow prohibitively compared to classical neural networks~\cite{kordzanganeh2023benchmarking, kuzmin2025tqml}. Part of the reason is the exponential scaling of classical data stored in qubits, but the gate-based decomposition compounds the problem, since each additional gate adds to the simulator runtime. Moreover, the circuit's ansatz (the architecture of gates applied to the device) constrains the data flow, and a given architecture offers no guarantees of optimality~\cite{bowles2024better, kandala2017hardware}.

    Ref.~\cite{mate2022beyond} works around this problem by abandoning gates and ans\"{a}tze altogether. M\'{a}t\'{e} et al.\ exploit the fact that all quantum operations, regardless of architecture or gate count, can be represented by a single unitary matrix. Unitary matrices, $U$, form an infinite family of square matrices whose dimension depends on the system size $N$. For a quantum computer, $N = 2^n$ where $n$ is the number of qubits. A unitary matrix of size $N$ can be completely described by $N^2 - 1$ real parameters, which are then optimized. Any ansatz is fully representable by such a matrix. Instead of training gate-by-gate, $N^2 - 1$ parameters are trained inside a skew-Hermitian matrix, which is then exponentiated to produce a unitary applied to the simulation as a single operation.

    In this paper, we build on that work with an ansatz-agnostic solution that avoids matrix exponentiation entirely. Section~\ref{sec: softU} shows how adding a regularization term to the loss function penalizes non-unitarity of the variational matrices, yielding objects close to unitary that we call soft-unitaries. Section~\ref{sec: alignment} describes how a variational quantum circuit can be aligned to a soft-unitary, and shows that this two-step indirect training can be faster than training VQCs with data directly. Section~\ref{sec: Experiments} demonstrates these advantages on two tasks: a supervised classification problem where the speed gain is most visible, and a reinforcement learning cartpole problem where soft-unitaries power a hybrid quantum-classical agent that outperforms a purely classical baseline. Finally, Section~\ref{sec: Conclusion} discusses the benefits and limitations of the technique.

%% ============================================================
%%  SOFT-UNITARIES
%% ============================================================
\section{Soft-Unitaries as a Method of Unitary Training}\label{sec: softU}

    Ansatz-agnostic unitaries can reach the entirety of the Hilbert space without any constraint from circuit architecture. In this way, they learn a function in a fixed amount of time independent of the number of gates. One drawback, however, is that maintaining unitarity during training is difficult. Changing a single element of an $N \times N$ unitary matrix will, in general, destroy the unitary property of the whole matrix. Since optimization methods change parameters, training an ansatz-agnostic unitary requires care.

    One route to guarantee unitarity is matrix exponentiation of a skew-Hermitian matrix, which can be constructed from a set of $N^2 - 1$ real parameters. This idea has a precedent in classical deep learning, where unitary recurrent neural networks enforce unitarity through structured matrix decompositions or Lie-algebra exponential maps~\cite{arjovsky2016unitary, lezcano2019cheap}. The parameters modified during training are then related to the final unitary only indirectly, through the exponentiation. For the large matrices encountered in quantum simulations, this is computationally expensive: not only for the forward pass, but even more so during backpropagation, where the optimizer must account for the effect of the matrix exponential on every parameter.

    This paper takes a different approach and abandons matrix exponentiation altogether. Instead, the elements of the unitary matrices are trained directly as parameters. Each matrix is initialized as a known unitary, and during training its unitarity is maintained by adding a regularization term to the loss function that penalizes deviations from unitarity:

\begin{equation}
    \mathcal{L}_{\text{total}} = \mathcal{L}_{\text{task}} + \mathcal{L}_{\text{unitary}}
\end{equation}

    where

\begin{equation}
     \mathcal{L}_{\text{unitary}} = \lambda \| U ^ \dagger U - \mathbb{I} \|
\end{equation}

    with the hyperparameter $\lambda$ controlling the strength of the regularization, $\mathbb{I}$ denoting the $2^n \times 2^n$ identity, and $\| A \|$ the matrix norm:

\begin{equation}
    \| A\| \equiv \sqrt{\text{tr}(A^\dagger A)}
\end{equation}

    Because the optimizer minimizes the total loss, the result is both a good solution to the machine learning task and close to unitary. We call these close-to-unitary objects \emph{soft-unitaries}.

    Training the matrix elements directly should be faster than exponentiating skew-Hermitian matrices~\cite{mate2022beyond} or routing data through gate-based operations. We note that regularization-based approaches have also been studied in the context of noise-induced training of quantum neural networks~\cite{kuzmin2025method}; the soft-unitary regularization differs in that it enforces a structural constraint (unitarity) rather than mimicking hardware noise. The limitation is that these matrices cannot be used at high qubit counts: they contain $\sim 4^n$ elements, so the memory and compute benefits of soft-unitaries are quickly outweighed by size constraints. Additionally, since the Lie algebra of soft-unitaries is guaranteed to be exponential in dimension, the loss landscape will exhibit barren plateaus~\cite{mcclean2018barren, cerezo2021cost, ragone2024lie}. More expressive ans\"{a}tze generally correlate with smaller gradient magnitudes~\cite{holmes2022connecting}, and the expressibility of soft-unitaries is maximal by construction~\cite{sim2019expressibility}.

    Despite these limitations, soft-unitaries serve as a useful research tool. They allow researchers to probe the expressive limit of an effectively infinitely deep variational layer and to understand what performance is achievable within a given Hilbert space before committing to a specific circuit architecture. Recent work on superposed parameterised circuits~\cite{patapovich2025superposed} explores related ideas by extending the expressibility of fixed-depth ans\"{a}tze through quantum superpositions of circuit configurations.

%% ============================================================
%%  CIRCUIT ALIGNMENT
%% ============================================================
\section{Circuit-Alignment Training}\label{sec: alignment}

    For the right number of qubits, soft-unitary training is fast and can fit complex functions well. But the soft-unitary matrices it produces must ultimately be placed on a quantum device. This section describes how we accomplish this using variational quantum compilers, in a process we call circuit alignment.

    Placing a given $2^n \times 2^n$ unitary matrix onto a quantum device is a well-studied problem. The Solovay-Kitaev theorem guarantees that any unitary operation can be approximated to arbitrary precision by a sequence of gates drawn from a fixed universal basis set~\cite{nielsen2010quantum}. The Quantum Shannon Decomposition breaks arbitrary unitaries into rotation and controlled-NOT gates in an efficient manner~\cite{shende2005synthesis}. However, soft-unitary training returns objects that are very close to unitary but not exact, which makes standard unitary compilation methods less reliable.

    Instead, the experiments in this paper use variational circuits to align as closely with the soft-unitary as possible. Variational quantum compilers~\cite{khatri2019quantum} are parameterized circuits similar to VQCs. Unlike quantum neural networks, which update parameters in response to new data, variational quantum compilers update parameters to match a constant target unitary $U_{\text{target}}$. The loss function for the alignment between the circuit and the target is

\begin{equation}\label{eq: alignment loss}
    \mathcal{L}_\text{alignment} = \frac{1}{2^n M}\sum_i^M \| U_{\text{target}, i } - U_{\text{circuit}, i }\|
\end{equation}

    In principle, the target can be any unitary, but the practical use case considered here sets $U_{\text{target}} = U_{\text{soft}}$. The depth required for accurate circuit alignment scales roughly as the number of trainable parameters in the target unitary, that is, $\sim 4^n$. This exponential scaling reinforces the point that soft-unitary training is restricted to few-qubit problems.

    The alignment loss (Equation~\ref{eq: alignment loss}) has three useful properties for training quantum neural networks. First, alignment is agnostic to how close $U_{\text{soft}}$ is to exact unitarity; the compiled circuit naturally finds the nearest unitary. Second, the end product is a variational quantum circuit, so fine-tuning remains possible if small-scale issues arise from the soft-unitary fit. Third, the method depends only on the target unitary and is completely independent of the original training data.

    Where soft-unitary training is fast because it is independent of the number of gates, circuit alignment is fast because it is independent of the size of the dataset. The target unitary is $2^n \times 2^n$ regardless of how many datapoints were used to train it.

    For $\mathbf{d}$ datapoints on a circuit with $\mathbf{g}$ gates, simulated training of a variational quantum circuit requires the data to pass through every gate, so the overall training time scales as

\begin{equation}
    \mathcal{O}(\mathbf{dg})
\end{equation}

    Soft-unitary training, by contrast, is $\mathcal{O}(1)$ in the number of gates $\mathbf{g}$, and circuit alignment is $\mathcal{O}(1)$ in the number of datapoints $\mathbf{d}$. Together, the two steps scale as

\begin{equation}
    \mathcal{O}(\mathbf{d}) + \mathcal{O}(\mathbf{g}),
\end{equation}

    which suggests that for low-qubit circuits, the combined cost of soft-unitary training followed by circuit alignment can be substantially lower than direct training of a variational quantum circuit.

%% ============================================================
%%  EXPERIMENTS
%% ============================================================
\section{Experiments}\label{sec: Experiments}

    We present two experiments that demonstrate the soft-unitary pipeline in different learning settings: a supervised classification task and a reinforcement learning task. The first focuses on training speed and accuracy relative to a standard VQC; the second shows that soft-unitaries integrate naturally into hybrid quantum-classical architectures and can outperform purely classical baselines.

\subsection{Top-hat function: supervised classification}\label{sec: tophat}

    We compare an ordinary VQC to the soft-unitary pipeline on a step-function classification task. This task is challenging for quantum models because they produce truncated Fourier series~\cite{schuld2021effect}, and linearly separable data can be difficult for quantum neural networks in general~\cite{bowles2024better}. Each model uses 5 qubits, and the one-dimensional features are preprocessed with an exponential $R_Z$ encoding following Ref.~\cite{kordzanganeh2022exponentially}.

\begin{figure}[H]
\centering
\includegraphics[width=\columnwidth]{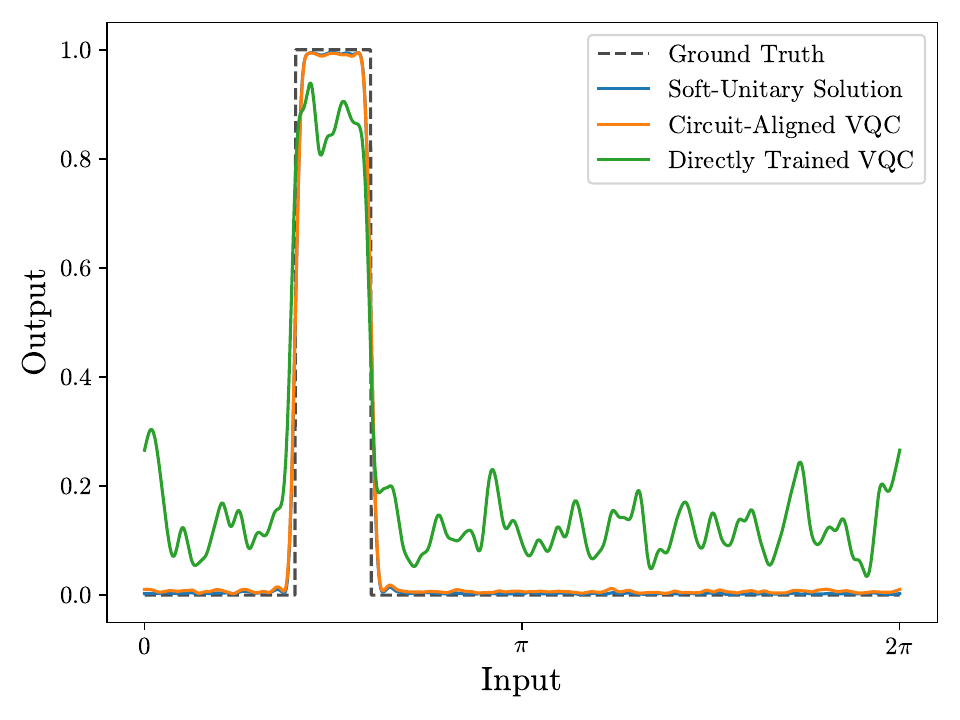}
\caption{Classification of a top-hat function comparing a directly trained VQC, the soft-unitary solution, and the circuit-aligned VQC. Outputs are rescaled from $[-1, 1]$ to $[0, 1]$ via a global shift and division by two. Both models use 5 qubits, 1000 datapoints, 200 epochs, 3 reuploading layers, and 4 variational layers. Circuit depth is 10 basic entangling layers per variational layer for the VQC and 69 basic entangling layers for the circuit-aligned soft-unitary.}
\label{fig:groundtruth_v_all}
\end{figure}

    Figure~\ref{fig:groundtruth_v_all} shows the results. Visually, the soft-unitary plus circuit alignment pipeline provides more accurate solutions.

\begin{figure}[H]
    \centering
    \includegraphics[width=\columnwidth]{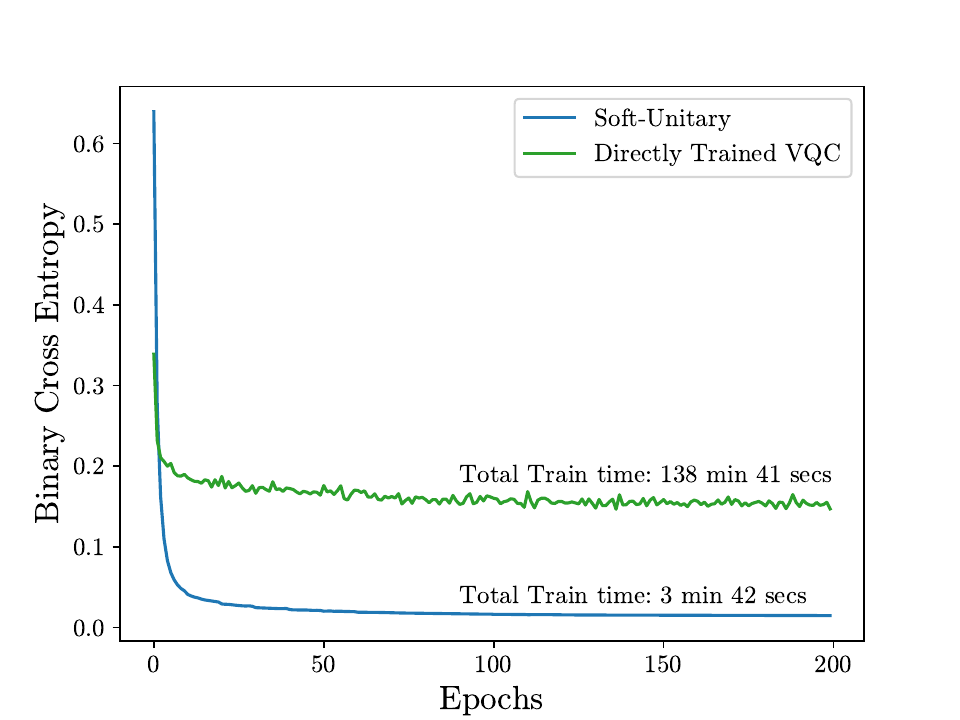}
    \caption{Training loss versus wall-clock time for the soft-unitary and the directly trained VQC. The soft-unitary loss decreases faster and reaches lower values because it is unconstrained by ansatz and independent of gate count.}
    \label{fig:time_to_train}
\end{figure}

    Binary cross-entropy losses for both training steps appear in Figure~\ref{fig:time_to_train}. The final soft-unitaries $U$ deviate from unitarity by $\| U^\dagger U - \mathbb{I} \| = 3 \times 10^{-4}$, close enough that the variational quantum circuit produced by alignment behaves as a well-trained model.

\begin{figure}[H]
    \centering
    \includegraphics[width=\columnwidth]{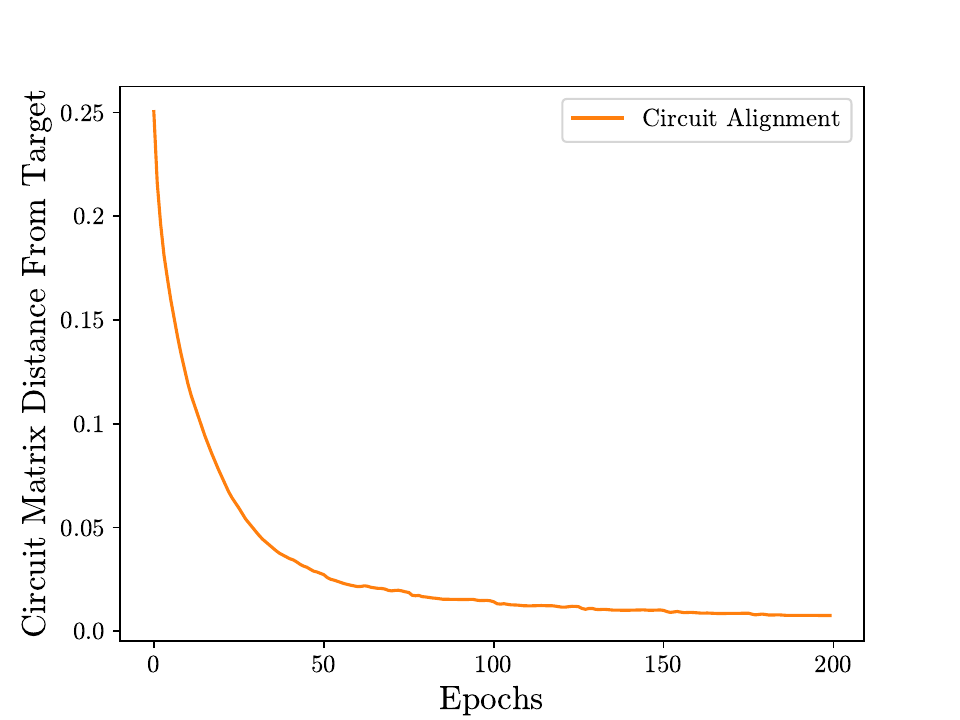}
    \caption{Circuit alignment loss, measured as the matrix norm $\| U_{\text{target}} - U_{\text{circuit}} \|$, versus epoch. Lower values indicate better agreement between the target soft-unitary and the compiled circuit.}
    \label{fig:circuit_align_loss}
\end{figure}

    The results also show a clear time advantage over direct VQC training, consistent with the complexity scaling presented in Section~\ref{sec: alignment}. Soft-unitary training ran for 200 epochs in 48 seconds. Circuit alignment (Figure~\ref{fig:circuit_align_loss}) then ran for another 200 epochs in 174 seconds, bringing the total training time for the end-product VQC to 3 minutes and 42 seconds.

\begin{figure}[H]
    \centering
    \includegraphics[width=\columnwidth]{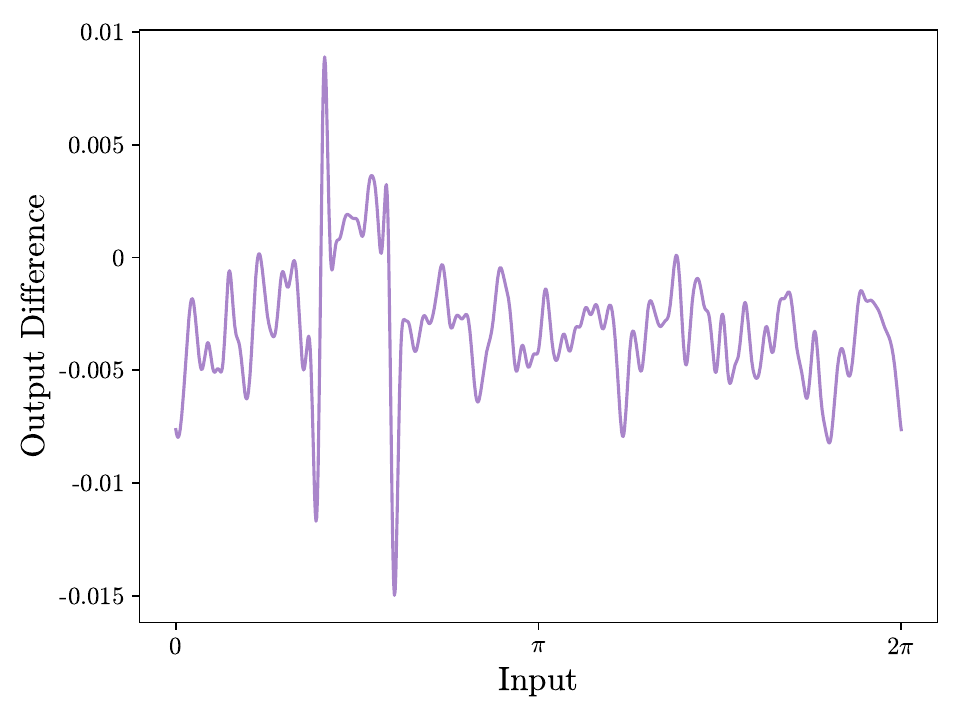}
    \caption{Difference in output values between the soft-unitary and the circuit-aligned model. The mean squared error between the two outputs is $9.6 \times 10^{-6}$.}
    \label{fig:soft_minus_aligned}
\end{figure}

    The difference between the aligned circuit and the soft-unitary is shown in Figure~\ref{fig:soft_minus_aligned}. For comparison, the directly trained VQC took roughly 42 seconds per epoch, or about 138 minutes total. The total time to produce a deployable model via the soft-unitary route is therefore less than 4 minutes versus over 2 hours for direct training.

    Ultimately, the soft-unitary pipeline performs so well because of its effective depth. The circuit-aligned model has as many parameters as the soft-unitary, which scales exponentially with qubit count. In this experiment, that meant 69 basic entangling layers for the circuit-aligned model, compared to only 10 for the directly trained VQC. Giving the VQC the same 69 layers would likely yield comparable accuracy, but at 257 seconds per epoch the estimated training time would be roughly 14 hours, over two orders of magnitude slower than the soft-unitary plus alignment approach.

\subsection{Cartpole: reinforcement learning with a hybrid architecture}\label{sec: cartpole}

\begin{figure}[H]
    \centering
    \includegraphics[width=\columnwidth]{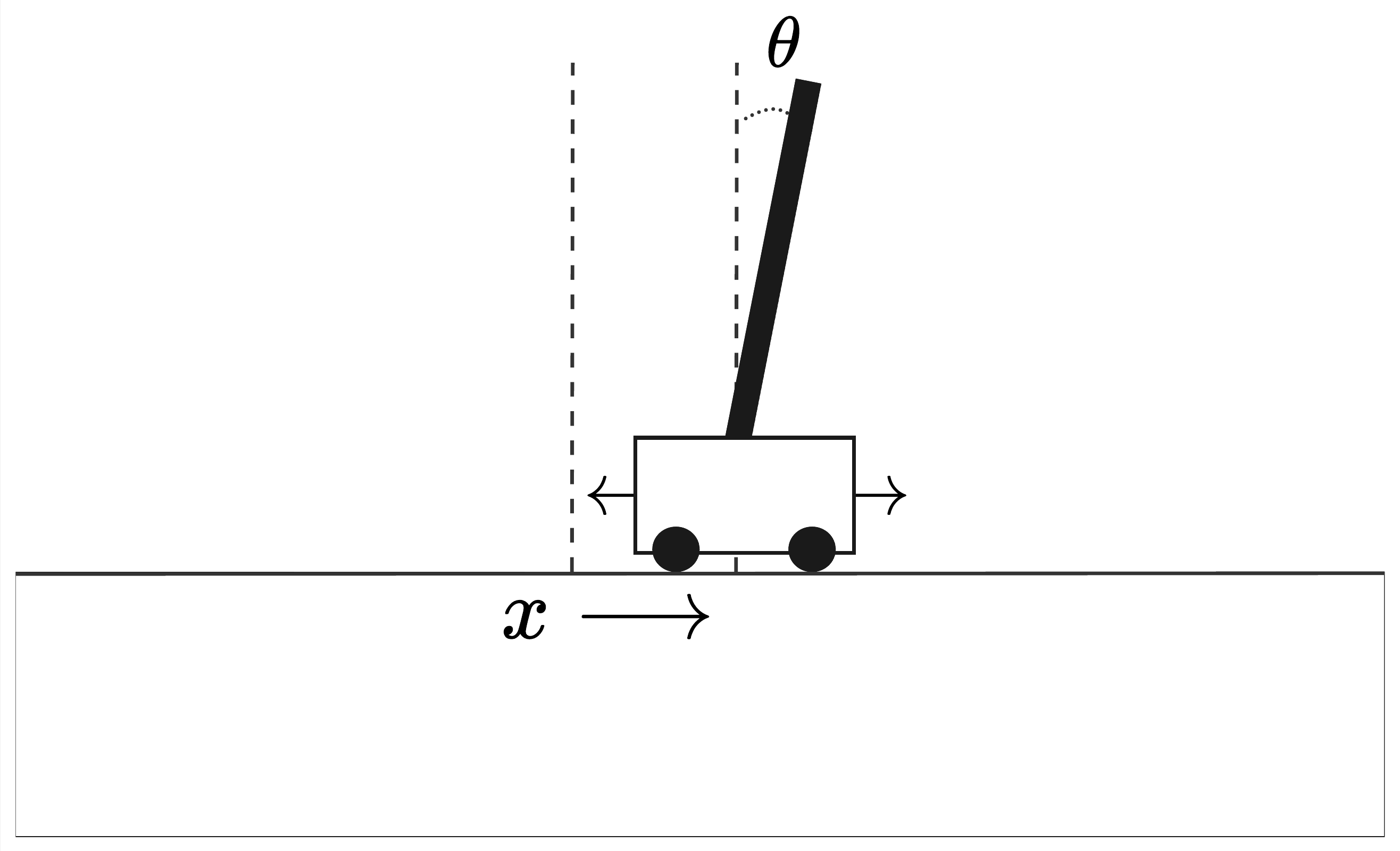}
    \caption{The cartpole problem. At each timestep the agent observes the cart's position, cart's velocity, pole angle, and pole angular velocity, and must decide whether to push the cart left or right to keep the pole upright for as long as possible.}
    \label{fig:cartpole_setup}
\end{figure}

    The cartpole problem is a standard reinforcement learning benchmark that simulates a two-dimensional robotics task. A pole is balanced on top of a cart that can only move left or right, and the agent must learn to stabilize the pole for as long as possible. The environment provides four features at each timestep: the cart's position, the cart's velocity, the pole's angle, and the pole's angular velocity. From these, the agent selects one of two actions (move left or move right). The simulation records how long the agent keeps the pole upright, with a maximum duration of 500 seconds. The setup is illustrated in Figure~\ref{fig:cartpole_setup}.

\begin{figure*}[!t]
\centering
\subfigure[3-layer classical multilayer perceptron]{
   \includegraphics[width=0.47\textwidth]{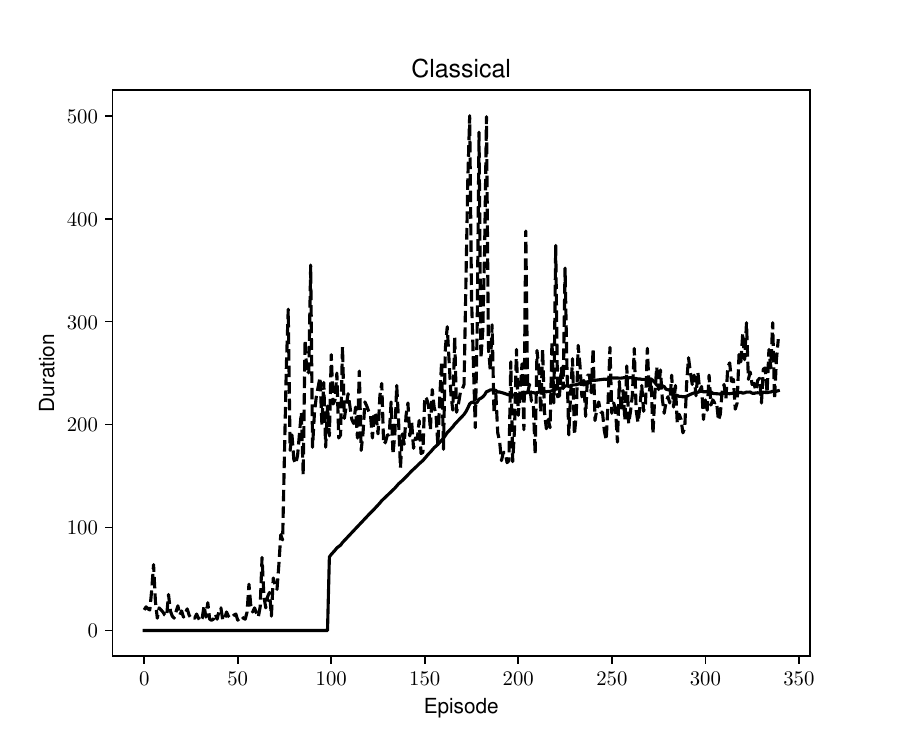}
   \label{fig:3layer_classical}
 }
\subfigure[2-layer hybrid neural network with soft-unitaries]{
   \includegraphics[width=0.47\textwidth]{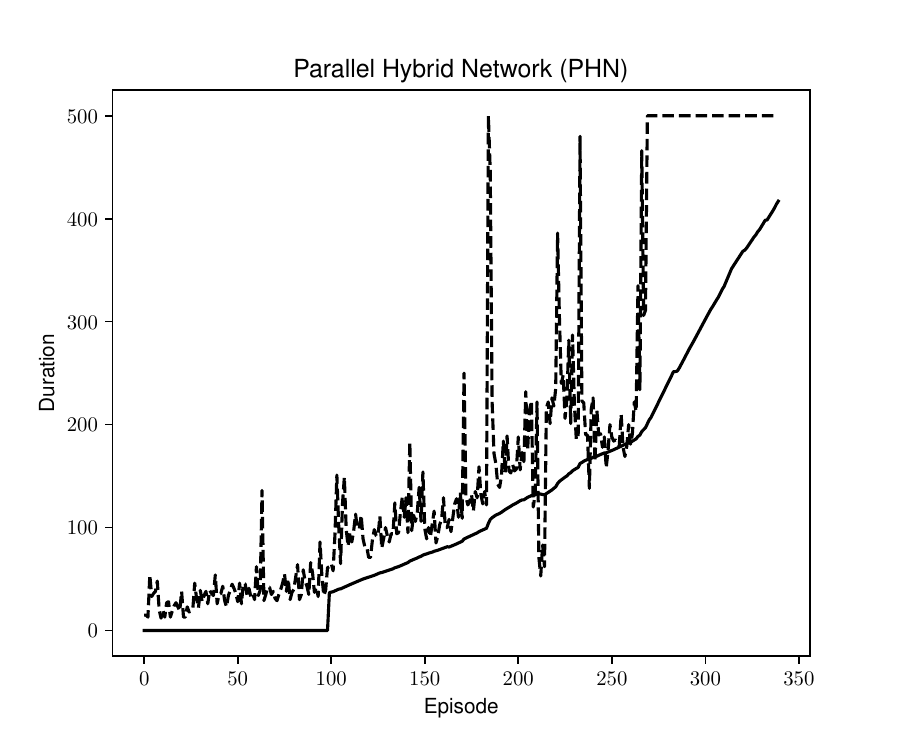}
   \label{fig:2layer_PHN}
 }
\caption{Comparison of the cartpole task between a purely classical and a hybrid quantum-classical neural network. Both architectures have roughly the same number of classical weights. Duration measures how long the agent keeps the pole upright (higher is better); the dotted line shows the raw duration and the solid line shows the running average. Over 340 episodes, the classical network reached a mean duration of $232.9$ while the hybrid model reached $417.0$. The last 71 episodes of the hybrid model were all maxed out at 500.}
\label{fig:cartpole_results}
\end{figure*}

    We train the agent using a deep-Q reinforcement learning protocol~\cite{mnih2013playing}. Quantum approaches to reinforcement learning have been explored with parameterized quantum policies~\cite{jerbi2021parametrized} and variational deep-Q networks~\cite{skolik2022quantum, lockwood2021playing}, but these works train gate-based circuits directly and thus face the same scaling constraints discussed in Section~\ref{sec: Intro}. Here we sidestep that overhead by using soft-unitaries. We compare two architectures: a 3-layer classical multilayer perceptron and a 2-layer parallel hybrid network (PHN)~\cite{kordzanganeh2023parallel, kurkin2025forecasting}. In the PHN, quantum and classical layers run in parallel. The quantum layers act as a Fourier neural operator~\cite{li2021fourier}, since data uploaded through rotation gates is encoded as a truncated Fourier series. This means the quantum branch captures large-scale structure while the classical branch resolves fine-grained details. The two networks are designed to have roughly the same number of classical weights, so that any performance difference can be attributed to the quantum layers.

    For the quantum layers we use 3 qubits. Since the quantum branch primarily captures long-wavelength Fourier terms, a small qubit count suffices. With rotation-gate encoding, however, the resulting circuits can be very deep. This would be prohibitive when training parameterized quantum circuits directly, but is irrelevant for soft-unitary training where the cost is independent of circuit depth. The difficulty of choosing an optimal ansatz for the quantum layers is also eliminated: the soft-unitary explores the full Hilbert space without any architectural constraint.

    Despite the fact that reinforcement learning does not relate to its loss function as directly as supervised learning does, a regularization strength of $\lambda = 1000$ proved sufficient to keep the soft-unitaries close to unitary throughout training. The results are shown in Figure~\ref{fig:cartpole_results}. The hybrid model with soft-unitaries consistently reaches the maximum duration faster than the purely classical network. Over 340 episodes, the classical network achieved a mean duration of $232.9$, while the hybrid model reached $417.0$. In the last 71 episodes, the hybrid model was maxed out at 500 in every run. The final unitary loss was $\mathcal{L}_{\text{unitary}} = 10^{-4}$, confirming that the soft-unitaries are nearly indistinguishable from true quantum operations. After circuit alignment, the alignment loss was $\mathcal{L}_{\text{alignment}} = 0.07$.

\medskip
%% ============================================================
%%  CONCLUSION
%% ============================================================
\section{Conclusion and Discussion}\label{sec: Conclusion}

    This paper presents a two-step training procedure that gives researchers a faster way to train deep, few-qubit circuits on large datasets. By separating quantum gates from data, the indirect training reduces wall-clock time by over two orders of magnitude compared to direct VQC training on the supervised benchmark. On the reinforcement learning task, soft-unitaries integrated into a hybrid architecture outperformed a purely classical network of comparable size, with the hybrid agent reaching a mean episode duration of $417.0$ versus $232.9$ for the classical baseline. As discussed in Sections~\ref{sec: softU} and~\ref{sec: alignment}, the technique has clear limitations: the exponential scaling of both soft-unitaries and circuit alignment, together with the guarantee of barren plateaus, restricts it to few-qubit problems. Within that regime, however, soft-unitaries can find high-quality quantum models in a fraction of the time required by direct training.

    We hope that soft-unitaries will help illuminate the limits of classical data embedded in small Hilbert spaces, and that the insights gained from this line of work will contribute to the broader development of quantum machine learning.

%% ============================================================
%%  REFERENCES
%% ============================================================

\bibliography{refs}

@article{peruzzo2014variational,
  title={A variational eigenvalue solver on a photonic quantum processor},
  author={Peruzzo, Alberto and McClean, Jarrod and Shadbolt, Peter and Yung, Man-Hong and Zhou, Xiao-Qi and Love, Peter J and Aspuru-Guzik, Al{\'a}n and O'brien, Jeremy L},
  journal={Nature Communications},
  volume={5},
  number={1},
  pages={4213},
  year={2014},
  publisher={Nature Publishing Group UK London}
}

@book{nielsen2010quantum,
  title={Quantum computation and quantum information},
  author={Nielsen, Michael A and Chuang, Isaac L},
  year={2010},
  publisher={Cambridge University Press}
}

@article{schuld2021effect,
  title={Effect of data encoding on the expressive power of variational quantum-machine-learning models},
  author={Schuld, Maria and Sweke, Ryan and Meyer, Johannes Jakob},
  journal={Physical Review A},
  volume={103},
  number={3},
  pages={032430},
  year={2021},
  publisher={APS}
}

@article{kordzanganeh2022exponentially,
  title={An exponentially-growing family of universal quantum circuits},
  author={Kordzanganeh, Mohammad and Sekatski, Pavel and Fedichkin, Leonid and Melnikov, Alexey},
  journal={Machine Learning: Science and Technology},
  volume={4},
  number={3},
  pages={035036},
  year={2023},
  doi={10.1088/2632-2153/ace757}
}

@article{mate2022beyond,
  title={Beyond {Ans\"{a}tze}: Learning quantum circuits as unitary operators},
  author={M{\'a}t{\'e}, B{\'a}lint and Saux, Bertrand Le and Henderson, Maxwell},
  journal={arXiv preprint arXiv:2203.00601},
  year={2022}
}

@article{sagingalieva2023hybrid,
  title={Hybrid quantum neural network for drug response prediction},
  author={Sagingalieva, Asel and Kordzanganeh, Mohammad and Kenbayev, Nurbolat and Kosichkina, Daria and Tomashuk, Tatiana and Melnikov, Alexey},
  journal={Cancers},
  volume={15},
  number={10},
  pages={2705},
  year={2023},
  publisher={MDPI}
}

@article{preskill2018quantum,
  title={Quantum computing in the {NISQ} era and beyond},
  author={Preskill, John},
  journal={Quantum},
  volume={2},
  pages={79},
  year={2018},
  publisher={Verein zur F{\"o}rderung des Open Access Publizierens in den Quantenwissenschaften}
}

@article{bowles2024better,
  title={Better than classical? The subtle art of benchmarking quantum machine learning models},
  author={Bowles, Joseph and Shahnawaz, Ahmed and Schuld, Maria},
  journal={arXiv preprint arXiv:2403.07059},
  year={2024}
}

@article{ragone2024lie,
  title={A {Lie} algebraic theory of barren plateaus for deep parameterized quantum circuits},
  author={Ragone, Michael and Bakalov, Bojko N and Sauvage, Fr{\'e}d{\'e}ric and Kemper, Alexander F and Marrero, Carlos Ortiz and Larocca, Martin and Cerezo, M},
  journal={Nature Communications},
  volume={15},
  pages={7172},
  year={2024},
  doi={10.1038/s41467-024-49909-3}
}

@article{mcclean2018barren,
  title={Barren plateaus in quantum neural network training landscapes},
  author={McClean, Jarrod R and Boixo, Sergio and Smelyanskiy, Vadim N and Babbush, Ryan and Neven, Hartmut},
  journal={Nature Communications},
  volume={9},
  number={1},
  pages={4812},
  year={2018},
  publisher={Nature Publishing Group UK London}
}

@inproceedings{shende2005synthesis,
  title={Synthesis of quantum logic circuits},
  author={Shende, Vivek V and Bullock, Stephen S and Markov, Igor L},
  booktitle={Proceedings of the 2005 Asia and South Pacific Design Automation Conference},
  pages={272--275},
  year={2005}
}

@article{mnih2013playing,
  title={Playing {Atari} with deep reinforcement learning},
  author={Mnih, Volodymyr and Kavukcuoglu, Koray and Silver, David and Graves, Alex and Antonoglou, Ioannis and Wierstra, Daan and Riedmiller, Martin},
  journal={arXiv preprint arXiv:1312.5602},
  year={2013}
}

@inproceedings{li2021fourier,
  title={{Fourier} neural operator for parametric partial differential equations},
  author={Li, Zongyi and Kovachki, Nikola and Azizzadenesheli, Kamyar and Liu, Burigede and Bhattacharya, Kaushik and Stuart, Andrew and Anandkumar, Anima},
  booktitle={International Conference on Learning Representations},
  year={2021}
}

@article{kordzanganeh2023parallel,
  title={Parallel hybrid networks: an interplay between quantum and classical neural networks},
  author={Kordzanganeh, Mo and Kosichkina, Daria and Melnikov, Alexey},
  journal={Intelligent Computing},
  volume={2},
  pages={0028},
  year={2023},
  publisher={AAAS},
  doi={10.34133/icomputing.0028}
}

@article{kordzanganeh2023benchmarking,
  title={Benchmarking simulated and physical quantum processing units using quantum and hybrid algorithms},
  author={Kordzanganeh, Mohammad and Buchberger, Markus and Kyriacou, Basil and Povolotskii, Maxim and Fischer, Wilhelm and Kurkin, Andrii and Somogyi, Wilfrid and Sagingalieva, Asel and Pflitsch, Markus and Melnikov, Alexey},
  journal={Advanced Quantum Technologies},
  volume={6},
  number={8},
  pages={2300043},
  year={2023},
  publisher={Wiley Online Library}
}

@article{cerezo2021variational,
  title={Variational quantum algorithms},
  author={Cerezo, M and Arrasmith, Andrew and Babbush, Ryan and Benjamin, Simon C and Endo, Suguru and Fujii, Keisuke and McClean, Jarrod R and Mitarai, Kosuke and Yuan, Xiao and Cincio, Lukasz and Coles, Patrick J},
  journal={Nature Reviews Physics},
  volume={3},
  number={9},
  pages={625--644},
  year={2021},
  publisher={Nature Publishing Group}
}

@article{bharti2022noisy,
  title={Noisy intermediate-scale quantum algorithms},
  author={Bharti, Kishor and Cervera-Lierta, Alba and Kyaw, Thi Ha and Haug, Tobias and Alperin-Lea, Sumner and Anand, Abhinav and Degroote, Matthias and Heimonen, Hermanni and Kottmann, Jakob S and Menke, Tim and others},
  journal={Reviews of Modern Physics},
  volume={94},
  number={1},
  pages={015004},
  year={2022},
  publisher={APS}
}

@article{kandala2017hardware,
  title={Hardware-efficient variational quantum eigensolver for small molecules and quantum magnets},
  author={Kandala, Abhinav and Mezzacapo, Antonio and Temme, Kristan and Takita, Maika and Brink, Markus and Chow, Jerry M and Gambetta, Jay M},
  journal={Nature},
  volume={549},
  number={7671},
  pages={242--246},
  year={2017},
  publisher={Nature Publishing Group}
}

@article{holmes2022connecting,
  title={Connecting ansatz expressibility to gradient magnitudes and barren plateaus},
  author={Holmes, Zo{\"e} and Sharma, Kunal and Cerezo, M and Coles, Patrick J},
  journal={PRX Quantum},
  volume={3},
  number={1},
  pages={010313},
  year={2022},
  publisher={APS}
}

@article{sim2019expressibility,
  title={Expressibility and entangling capability of parameterized quantum circuits for hybrid quantum-classical algorithms},
  author={Sim, Sukin and Johnson, Peter D and Aspuru-Guzik, Al{\'a}n},
  journal={Advanced Quantum Technologies},
  volume={2},
  number={12},
  pages={1900070},
  year={2019},
  publisher={Wiley Online Library}
}

@article{cerezo2021cost,
  title={Cost function dependent barren plateaus in shallow parametrized quantum circuits},
  author={Cerezo, M and Sone, Akira and Volkoff, Tyler and Cincio, Lukasz and Coles, Patrick J},
  journal={Nature Communications},
  volume={12},
  number={1},
  pages={1791},
  year={2021},
  publisher={Nature Publishing Group UK London}
}

@article{khatri2019quantum,
  title={Quantum-assisted quantum compiling},
  author={Khatri, Sumeet and LaRose, Ryan and Poremba, Alexander and Cincio, Lukasz and Sornborger, Andrew T and Coles, Patrick J},
  journal={Quantum},
  volume={3},
  pages={140},
  year={2019},
  publisher={Verein zur F{\"o}rderung des Open Access Publizierens in den Quantenwissenschaften}
}

@inproceedings{arjovsky2016unitary,
  title={Unitary evolution recurrent neural networks},
  author={Arjovsky, Martin and Shah, Amar and Bengio, Yoshua},
  booktitle={International Conference on Machine Learning},
  pages={1120--1128},
  year={2016},
  organization={PMLR}
}

@inproceedings{lezcano2019cheap,
  title={Cheap orthogonal constraints in neural networks: A simple parametrization of the orthogonal and unitary group},
  author={Lezcano-Casado, Mario and Mart{\'\i}nez-Rubio, David},
  booktitle={International Conference on Machine Learning},
  pages={3794--3803},
  year={2019},
  organization={PMLR}
}

@article{mitarai2018quantum,
  title={Quantum circuit learning},
  author={Mitarai, Kosuke and Negoro, Makoto and Kitagawa, Masahiro and Fujii, Keisuke},
  journal={Physical Review A},
  volume={98},
  number={3},
  pages={032309},
  year={2018},
  publisher={APS}
}

@article{benedetti2019parameterized,
  title={Parameterized quantum circuits as machine learning models},
  author={Benedetti, Marcello and Lloyd, Erika and Sack, Stefan and Fiorentini, Mattia},
  journal={Quantum Science and Technology},
  volume={4},
  number={4},
  pages={043001},
  year={2019},
  publisher={IOP Publishing}
}

@inproceedings{jerbi2021parametrized,
  title={Parametrized quantum policies for reinforcement learning},
  author={Jerbi, Sofiene and Gyurik, Casper and Marshall, Simon C and Briegel, Hans J and Dunjko, Vedran},
  booktitle={Advances in Neural Information Processing Systems},
  volume={34},
  pages={28362--28375},
  year={2021}
}

@article{skolik2022quantum,
  title={Quantum agents in the {Gym}: a variational quantum algorithm for deep {Q}-learning},
  author={Skolik, Andrea and Jerbi, Sofiene and Dunjko, Vedran},
  journal={Quantum},
  volume={6},
  pages={720},
  year={2022},
  publisher={Verein zur F{\"o}rderung des Open Access Publizierens in den Quantenwissenschaften}
}

@inproceedings{lockwood2021playing,
  title={Playing {Atari} with hybrid quantum-classical reinforcement learning},
  author={Lockwood, Owen and Si, Ming},
  booktitle={NeurIPS 2020 Workshop on Pre-registration in Machine Learning},
  pages={285--301},
  year={2021},
  organization={PMLR}
}

@article{qml_review_2023,
  title={Quantum machine learning: from physics to software engineering},
  author={Melnikov, Alexey and Kordzanganeh, Mohammad and Alodjants, Alexander and Lee, Ray-Kuang},
  journal={Advances in Physics: X},
  volume={8},
  number={1},
  pages={2165452},
  year={2023},
  publisher={Taylor \& Francis},
  doi={10.1080/23746149.2023.2165452}
}

@article{kuzmin2025tqml,
  title={{TQml} Simulator: optimized simulation of quantum machine learning},
  author={Kuzmin, Viacheslav and Kyriacou, Basil and Protasevich, Tatjana and Papierz, Mateusz and Kordzanganeh, Mo and Melnikov, Alexey},
  journal={arXiv preprint arXiv:2506.04891},
  year={2025}
}

@article{kuzmin2025method,
  title={Method for noise-induced regularization in quantum neural networks},
  author={Kuzmin, Viacheslav and Somogyi, Wilfrid and Pankovets, Ekaterina and Melnikov, Alexey},
  journal={Advanced Quantum Technologies},
  volume={8},
  number={12},
  pages={e00603},
  year={2025},
  publisher={Wiley Online Library},
  doi={10.1002/qute.202400603}
}

@article{patapovich2025superposed,
  title={Superposed parameterised quantum circuits},
  author={Patapovich, Viktoria and Periyasamy, Maniraman and Kordzanganeh, Mo and Melnikov, Alexey},
  journal={arXiv preprint arXiv:2506.08749},
  year={2025}
}

@article{kurkin2025forecasting,
  title={Forecasting steam mass flow in power plants using the parallel hybrid network},
  author={Kurkin, Andrii and Hegemann, Jonas and Kordzanganeh, Mo and Melnikov, Alexey},
  journal={Engineering Applications of Artificial Intelligence},
  volume={160},
  pages={111912},
  year={2025},
  doi={10.1016/j.engappai.2025.111912}
}

\end{document}